# Simultaneous Raman based power combining and wavelength conversion of high-power fiber lasers

**SANTOSH APARANJI[†], V. BALASWAMY[†], S. ARUN, AND V. R. SUPRADEEPA[*]**

*Centre for Nano Science and Engineering, Indian Institute of Science, Bangalore 560012, India*
*\*supradeepa@iisc.ac.in*
*[†] These authors contributed equally to this paper*

**Abstract:** We present a technique for simultaneous power-combining and wavelength-conversion of multiple fiber lasers into a single, longer wavelength in a different band through Raman-based, nonlinear power combining. We illustrate this by power combining of two independent Ytterbium lasers into a single wavelength around 1.5micron with high output powers of upto 99W. A high conversion efficiency of ~64% of the quantum limited efficiency and a high level of wavelength conversion with >85% of the output power in the final wavelength is demonstrated. The proposed method enables power-scaling in various wavelength bands where conventional fiber lasers are unavailable or limited in power.

**References and links**

1. D. J. Richardson, J. Nilsson and W. A. Clarkson, "High power fiber lasers: current status and future perspectives [Invited]," J. Opt. Soc. Am. B **27**, B63-B92 (2010).
2. C. Jauregui, J. Limpert, and A. Tunnermann, "High power fibre lasers," Nat. Photonics **7**, 861-867 (2013).
3. H. M. Pask, R. J. Carman, D. C. Hanna, A. C. Tropper, C. J. Mackechnie, P. R. Barber, and J. M. Dawes, "Ytterbium-doped silica fiber lasers: versatile sources for the 1-1.2 μm region," *IEEE J. Sel. Top. Quant. Electron.* **1,** 2-13 (1995).
4. J. W. Dawson, M. J. Messerly, R. J. Beach, M. Y. Shverdin, E. A. Stappaerts, A. K. Sridharan, P. H. Pax, J. E. Heebner, C. W. Siders, and C.P.J. Barty, "Analysis of the scalability of diffraction-limited fiber lasers and amplifiers to high average power," Opt. Express **16**, 13240-13266 (2008).
5. J. Zhu, P. Zhou, Y. Ma, X. Xu, and Z. Liu, "Power scaling analysis of tandem-pumped Yb-doped fiber lasers and amplifiers," Opt. Express **19**, 18645-18654 (2011).
6. J. Zhu, P. Zhou, X. Wang, X. Xu, and Z. Liu, "Analysis of maximum extractable power of single-frequency $Yb^{3+}$-doped phosphate fiber sources," *IEEE J. Sel. Top. Quant.*, **48**, 480-484 (2012).
7. V. R. Supradeepa, Y. Feng, and J. W. Nicholson, "Raman fiber lasers," J. Optics **19**, 023001 (2017).
8. V. Kuhn, D. Kracht, J. Neumann, and P. Wessels, "Er-doped photonic crystal fiber amplifier with 70 W of output power," Opt. Lett. **36**, 3030–3032 (2011).
9. Y. Jeong, S. Yoo, C. A. Codemard, J. Nilsson, J. K. Sahu, D. N. Payne, R. Horley, P. W. Turner, L. Hickey, A. Harker, and M. Lovelady, "Erbium:Ytterbium co-doped large core fiber laser with 297-W continuous-wave output power," *IEEE J. Sel. Top. Quant. Electron.* **13**, 573-579 (2007).
10. V. R. Supradeepa, J. W. Nicholson, C. E. Headley, M. F. Yan, B. Palsdottir, and D. Jakobsen, "A high efficiency architecture for cascaded Raman fiber lasers," Opt. Express **21**, 7148-7155 (2013).
11. V. R. Supradeepa and J. W. Nicholson, "Power scaling of high-efficiency 1.5 μm cascaded Raman fiber lasers," Opt. Lett. **38**, 2538-2541 (2013).
12. V. R. Supradeepa, J. W. Nicholson, C. E. Headley, Y. Lee, B. Palsdottir, and D. Jakobsen, "Cascaded Raman fiber laser faser at 1480nm with output power of 104W," no. 8237–48, SPIE photonics west 2012.
13. S. Aparanji, V. Balaswamy, and V. R. Supradeepa, "On the stability of Raman fiber lasers," in *13th International Conference on Fiber Optics and Photonics 2016,* OSA Technical Digest (Optical Society of America, 2016), paper Th3A.20. (2016)
14. C. X. Yu, S. J. Augst, S. M. Redmond, K. C. Goldizen, D. V. Murphy, A. Sanchez, and T. Y. Fan, "Coherent combining of a 4 kW, eight-element fiber amplifier array," Opt. Lett. **36**, 2686–2688 (2011).
15. S. J. Augst, A. K. Goyal, R. L. Aggarwal, T. Y. Fan, and A. Sanchez, "Wavelength beam combining of ytterbium fiber lasers," Opt. Lett**. 28**, 331–333 (2003).
16. S. A. Babin, E. A. Zlobina, S. I. Kablukov, and E. V. Podivilov, "High-order random Raman lasing in a PM fiber with ultimate efficiency and narrow bandwidth," Sci. Rep. **6**, 22625 (2016).
17. L. Zhang, H. Jiang, X. Yang, W. Pan, and Y. Feng, "Ultra-wide wavelength tuning of a cascaded Raman random fiber laser," Opt. Lett. **41**, 215-218 (2016).

18. S. Arun, V. Balaswamy, S. Aparanji and V. R. Supradeepa, "High-power, grating-free, cascaded Raman fiber lasers," in *Conference on Lasers and Electro-optics (CLEO) Europe-EQEC 2017,* CJ-2.4 (2017)
19. V. Balaswamy, S. Arun, Santosh Aparanji, Vishal Choudhury and V. R. Supradeepa, "High Power, Fixed and Tunable Wavelength, Grating-Free Cascaded Raman Fiber Lasers," arXiv:1711.10966 [physics.optics] (2017)
20. V. Balaswamy, S. Aparanji, G. Chayran, and V. R. R. Supradeepa, "Tunable Wavelength, Tunable Linewidth, High Power Ytterbium Doped Fiber Laser," in 13th International Conference on Fiber Optics and Photonics, OSA Technical Digest (online) (Optical Society of America, 2016), paper Tu3E.4. (2016)

## 1. Introduction

Fiber lasers have seen tremendous advance in recent years, finding applications in various defence, scientific and industrial sectors, owing to their superior power scaling abilities as compared to their solid-state and gas laser counterparts [1-3]. The attractiveness of fiber lasers stems from the unique qualities of waveguiding, large surface-area to volume ratio and the ability to be doped with high concentration of rare-earth dopants. This results in better beam quality and, better thermal management [1, 2]. Over 10 kW from a single-mode fiber laser has been demonstrated [1]. Leading this upward trend are Ytterbium (Yb) doped fiber lasers (YDFLs), operating in the 1.05-1.1micron wavelength region. This is due to several favorable properties like low quantum defect and ability to support higher doping concentrations. The obtained powers are quite close to the maximum theoretical extractable power from these lasers [4-6]. Fiber lasers with other rare-earth dopants like Erbium, Thulium, Holmium, Bismuth exist, but these lasers don't provide as much power scaling as YDFLs at their respective emission wavelengths [1]. Even including the other fiber lasers, power scaling has been confined to certain specific wavelength bands with large white spaces in between, where there exists no mature conventional rare-earth doped fiber laser technology to efficiently emit high powers. However, certain desirable attributes of eye-safety and atmospheric transparency are found wanting in the Yb emission band, but exist in other wavelength bands such as the 1.5micron band [7]. For the 1.5micron band specifically, Erbium-doped or Erbium-Ytterbium co-doped fiber lasers can be utilized, but power-scaling in these lasers is limited due to high thermal load and/or reduced beam quality with Erbium doped lasers [8] and also parasitic lasing in case of Erbium-Ytterbium co-doped lasers [9].

One attractive, power-scalable solution to achieve wavelength agility are cascaded Raman fiber lasers, which can provide high optical powers in a variety of wavelength bands. In essence, such Raman fiber lasers take as input a high-power laser in one of the accessible wavelength bands and use the process of stimulated Raman scattering to get to the otherwise inaccessible wavelengths through a series of Stokes shifts. Several architectures have been proposed to efficiently convert an input Ytterbium laser to the required wavelength through cascaded Raman conversion [7, 10-12]. Conventionally, these cascaded Raman lasers consist of a single mode fiber with smaller core size to enhance the Raman nonlinear coefficient, referred to as Raman fiber in nested cavities at the Stokes wavelengths of the input pump. Due to the losses in the resonator, such an architecture provided a maximum output power of ~ 100W with 45% conversion efficiency [12]. This was improved upon by a new, single pass, seeded architecture which used a low power conventional cascaded Raman resonator to seed all the intermediate wavelengths into the Raman fiber to enhance preferential forward scattering in the Raman fiber. This architecture achieved a substantially higher efficiency of 64%[10-11].

In all these architectures, there exists a single, high power Yb laser at the input, acting as the pump for the cascaded Raman conversion. One intuitive and seemingly straightforward way to scale the output power of such a Raman laser would be to scale the power of the input pump source. However, as we increase the input power, there is an onset of temporal oscillations and deleterious instabilities in the cascaded Raman laser due to coupling between the cascaded Raman converter and the rare-earth doped fiber laser [7, 13]. Also, scaling the power by scaling the input pump power leads to additional stress on the components. Moreover, because of the lack of redundancy in the input, pump failure would inevitably lead to system failure.

Therefore, it would be expedient to think of a method to overcome these limitations imposed by the usage of a single pump source.

In this work, we address the question of power-scaling with multiple lower-power laser modules. Specifically, is it possible to have multiple lower-power laser modules, all possibly operating at different wavelengths, but in the same band, and by passing them through a cascaded Raman converter, hope to achieve higher stability, simultaneous power-scaling and wavelength conversion to a single lasing line at any different wavelength band?

Conventionally, power combining to obtain a single lasing wavelength utilizes complex architectures to control the phase or wavelength of each of the input lasers, such as coherent and spectral power combining [14, 15]. In this work, we propose a technique for simultaneous power-combining and wavelength-conversion of multiple Ytterbium-doped fiber lasers into a single wavelength in a different band through Raman-based, nonlinear power combining. We illustrate this by power combining two independent Ytterbium lasers into a single wavelength around 1.5micron with high output power of ~99W.

## 2. Experimental setup

The conceptual schematic of such a nonlinear power combining technique is illustrated in Fig. 1, where the principle is illustrated for the case of two independently controlled Ytterbium-doped fiber lasers operating at two different wavelengths in the Yb emission band. This conceptual analysis is equally applicable to the more general case of multiple ytterbium lasers as input.

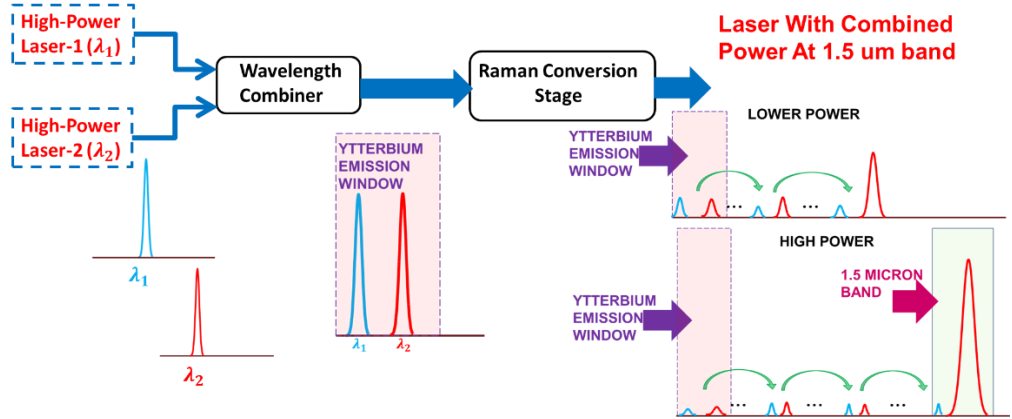

Fig. 1. Schematic illustrating simultaneous power combining and wavelength conversion.

Suppose we had only one Yb laser operating at the wavelength $\lambda_1$ in the above schematic and we feed it into the cascaded Raman converter module. It would end up at some particular Stokes-shifted output, say $\lambda_{S1}$, corresponding to the input wavelength $\lambda_1$. If instead, we had the other Yb-doped laser, operating at the wavelength $\lambda_2$, acting alone it would produce at the output a different wavelength than $\lambda_{S1}$, say $\lambda_{S2}$, corresponding to its particular Stokes-shifted wavelength. If we now wavelength-combine these two independent and uncorrelated Yb-doped fiber lasers and feed them into the Raman conversion module, one might intuitively think that one would observe a dual wavelength output laser where both $\lambda_{S1}$ and $\lambda_{S2}$ are present at the output. However, this is surprisingly not the case for what is present at the output of the cascaded Raman converter module are not two lasing wavelengths but a single lasing line corresponding to either $\lambda_{S1}$ or $\lambda_{S2}$. This happens because of the asymmetric shape of the Raman gain spectrum in Germania-doped silica fibers and the cross-coupling between the Stokes lines of the two lasers within the Raman converter. This results in one of the Stokes lines dominating over the other as shown in fig 1 and at some point there would be a complete transfer of power from the Stokes line of one of the laser into the other. This results in power-combining of the two lasers and simultaneous wavelength-conversion. In the circumstance that the Raman gain profile were symmetric around the peak, we anticipate that the Raman gain seen by a longer wavelength from a shorter wavelength is comparable to the loss seen by the longer wavelength to the wavelength which is two Stokes shifts of the shorter wavelength. In this case, we expect greater continued separation of the Stokes components of the individual lasers rather than merging of the two.

The above approach of using multiple, lower power sources is anticipated to enhance stability. Raman instability is a two-step process related to both the amount of spurious backward light moving from the Raman convertor to the input pump and the amount of amplification this light sees in the rare-earth doped fiber laser [13]. By moving to multiple pumps, despite the backward light from the Raman convertor being of comparable magnitude, its relative impact is reduced due to individual powers and amplification in the multiple pumps being smaller in this case. We anticipate that the enhancement in stability is because, the instability is decided by whether the power of the individual modules is beyond a threshold power while the total output power from the Raman laser is decided by the total power of all the modules.

In order to implement this conceptual schematic, three basic functionalities are necessary. Firstly, it would need a wavelength-independent feedback mechanism in order to be color-blind to the input wavelengths. Irrespective of the input wavelength, the cascaded Raman converter will be able to produce a Stokes-shifted output. Secondly a mechanism is needed to terminate the cascaded Raman conversion at the desired wavelength band. And lastly, a wavelength-multiplexer is required at the input in order to combine the input wavelengths prior to feeding the multiplexed input to the cascaded Raman converter.

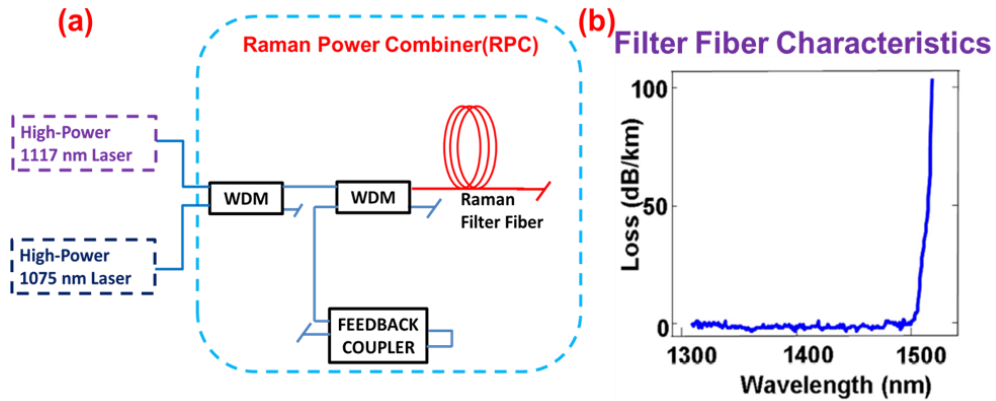

Fig. 2. (a) Architecture implementing the nonlinear Raman-based power combining. (b) Loss characteristics of the Raman filter fiber (from [11]).

The architecture that implements this nonlinear, Raman-based power combining technique is illustrated in Fig. 2(a), and this architecture incorporates all the necessary functionalities outlined above. The first Wavelength Division Multiplexer(WDM) multiplexes the two input lasers following which it is fed into the cascaded Raman convertor.

In recent times, the concept of a distributed feedback Raman fiber laser has been proposed [16, 17], and this concept has been the key enabler for developing a fixed and tunable wavelength, grating-free architecture of a high power Raman laser [17-19], where the Raman fiber is in a semi-open cavity with feedback provided by the 4% Fresnel reflection (due to the presence of a flat cleave) of the distributed backscatter, which is independent of the input wavelength. Such an architecture has demonstrated a color-blind mechanism to convert any input wavelength to high order Stokes wavelengths thereby eliminating the use of fixed grating sets and providing a wavelength-independent feedback mechanism. This grating-free architecture has been incorporated into the nonlinear Raman-based power combining architecture in Fig. 2(a) in order to provide a wavelength-independent feedback. The feedback coupler is used in lieu of a flat-cleave in previous work to enhance the amount of feedback. A 80/20 fused fiber coupler with terminated ends is utilized which provides a feedback of ~36%, reasonably higher that the 4% feedback provided by a flat-cleave. This coupler's feedback serves the purpose of sending back the distributed feedback at all the intermediate Stokes generated in the Raman filter fiber. This is essential for the preferential forward scattering in order to make the process of Stimulated Raman scattering more efficient by the process of seeding with the intermediate Stokes wavelengths.

One of the limitations with broadband feedback is termination of the cascade. As input power is increased, the output wavelength is further converted to the next Stokes component, compromising the efficiency and spectral purity. This previously limited the achieved power [16, 17]. In our case, we use a secondary mechanism to terminate the cascade at the required wavelength band through the use of a fiber which has a cut-off (high loss) at the designed long wavelength side [12]. This ensures that the Raman cascade terminates at the desired wavelength band. Such fibers, based often on a W-shaped index profile, are referred to as Raman filter fibers. The Raman filter fiber used in the current experiments has a cutoff at 1500nm resulting in wavelength conversion to the 1.5micron band. Its loss characteristics is shown in Fig. 2(b) [11] and the gain profile can be found in [7]. By suitable choice of the cut-off wavelength of the Raman filter fiber, wavelength conversion to any band in the transmission window of silica optical fibers can be envisioned.

In our experiment, we have two high power lasers of the 100W class, both operating in the Ytterbium band, at 1117 nm and 1075 nm. It is to be noted that these two lasers are not separated by one Stokes shift in order to distinguish it from the special case where the two lasers are power combined from within the same Raman cascade series, and hence this is a more general case. The optical power emanating from the 1117 nm laser is 98.1 W and that from the 1075 nm laser is 85.3 W, both at full power. After being wavelength-multiplexed by the high power, fused fiber WDM, they are fed to 160 m of Raman filter fiber with an effective area of 12sq-micron through another WDM, whose other input port is spliced to a looped coupler to provide feedback. In essence, this stage forms the core of the nonlinear Raman Power Combiner (RPC) where the simultaneous wavelength conversion and power combining occurs. In this work, we used a fused WDM which can handle upto 500W of total input power, but in reality, for further power scaling, especially when multiple input lasers are used, free-space beam combiners can also be used instead.

## 3. Results

Figure 3(a) shows the final spectrum at the output of the Raman filter fiber for the full coupled input power of 183.4W from the 2 independent lasers. The total output power is 102 W and an impressive 87.1W of this total optical power resides in the 1.5 micron band. This corresponds to a fraction of 85% of the total output power in the 1.5 micron band.

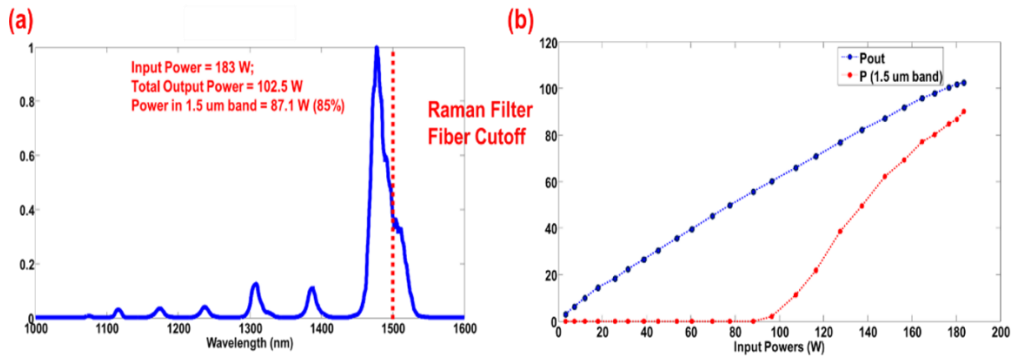

Fig. 3. (a) Spectrum at the output of the Raman filter fiber for a combined coupled input power of 183.4 W. (b) Power plot showing the evolution of the total output power and the power in the 1.5 micron band as a function of the combined input power.

Figure 3(b) shows the power plot of the total recorded output power from the RPC(blue) and the power in the 1.5 micron band(red) as a function of the combined input power of the two lasers. From this plot it can be seen that as the input powers are increased, there is a greater degree of conversion to the 1.5 micron band. The effective quantum limited conversion efficiency in this experiment is ~73% and at full power we obtain a conversion efficiency of 47%, which corresponds to ~ 64% of the quantum limited efficiency.

Here we have demonstrated a proof-of-concept experiment with two lasers. But in principle, this concept of nonlinear Raman-based power combining can be extended to the case with more than two input lasers, at various wavelengths. As an initial step we have performed the experiment with the two input lasers at different wavelengths as illustrated in Figs. 4(a)-4(c) to investigate the power combinability and as seen in the figure, that they are still power combinable. Based on the availability of lasers, one is fixed at 1117nm, while the other is a tunable, high power Yb laser source [20]. Figure 4(b) shows that the two input lasers are operating at the wavelengths 1117nm and 1070nm, both at the 100W class, and at a combined coupled input power of 193.4W, we obtained 94.3W of output power in the 1.5 micron band for a total output of 107.8W, corresponding to a fraction >87% in the desired band.

Figure 4(c) illustrates the special case where the two input lasers operate at the wavelengths of 1117 nm and 1064 nm, being separated by one Stokes shift with respect to each other. In this case for a combined input coupled power of 201.1 W, we obtained ~ 100W at the 1.5 micron band for a total output power of 112.4 W, which corresponds to ~ 88% of the total output power vesting in the 1.5 micron band. In this special case, we anticipate that the 1064nm initially amplifies and gets converted to 1117nm resulting in the merging of the two lines with the first Stokes shift followed by cascaded conversion to the 1.5micron region effectively as a single source at 1117nm. Because of this reason, the conversion efficiency is also the highest in this case. Figs. 4(d)-4(e) show the power plots as a function of the combined coupled input optical power, and it can be seen in both cases that there is a greater degree of conversion to the 1.5 micron band with increasing input power. Here, we notice that the amount of coupled power into the RPC is dependent on the input wavelengths. This is due to the use of a fused WDM specified for use with 1064/1117nm. Suitable WDM design and selection to ensure low-loss coupling of all input sources would be essential as we scale the number of input lasers.

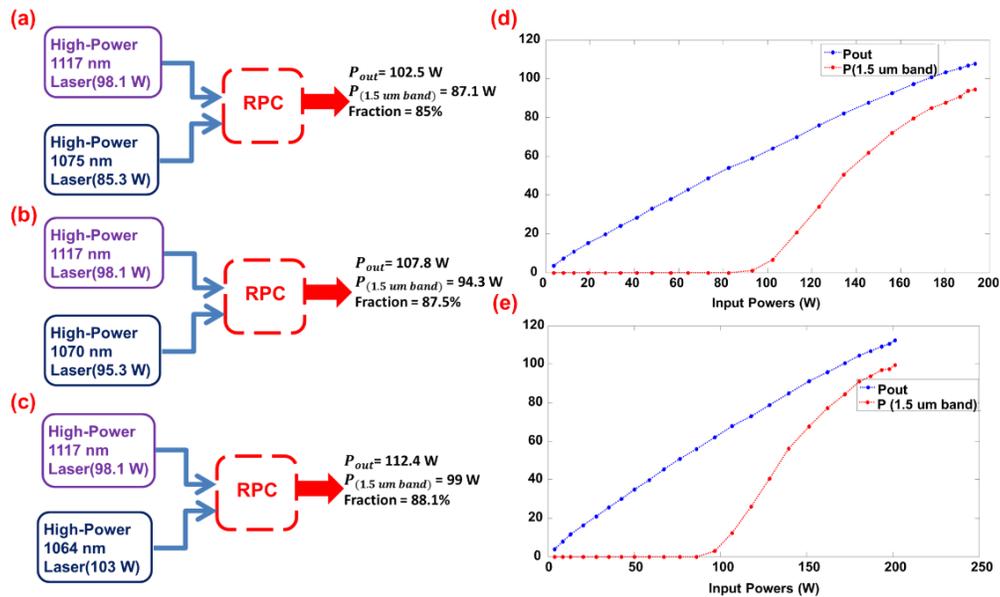

Fig. 4. (a), (b), (c) Schematic representation of the results of the power combining for the case of three different input wavelength combinations. (d), (e) Power plot showing the evolution of total output power and the power in the 1.5 micron band with the variable input laser operating at 1070 nm and 1064 nm respectively.

In the future, we will be investigating the efficiency of this architecture as the wavelength spacing between the individual input pump lasers decrease. In particular, we are interested to look at the dynamics of the multiple input lasers merging into a single line as a function of the wavelength spacing between them and the number of Stokes shifts utilized.

In summary, we have demonstrated a high power 1.5micron laser obtained through the simultaneous power combining and wavelength conversion of two independently controlled Ytterbium doped fiber lasers operating at different wavelengths. This approach can be scaled to multiple Ytterbium lasers and can be wavelength converted to any desired band in the transparency region of the optical fiber used. Such a system promises to be an attractive solution for scalable power at various wavelength bands and applications enabled by it.

## Funding